\documentclass[10pt,letterpaper]{article}
\usepackage{opex3}
\usepackage{amsmath}

\begin{document}
%% NOTE: TITLE PAGE & TOC NOT USED FOR MANUSCRIPT SUBMISSIONS %%
%\title{Experimental generation of amplitude squeezed vectorial modes}
%
%\vskip4pc
%
%\tableofcontents
%\clearpage
%% NO TITLE PAGE FOR OPEX SUBMISSIONS %%
%
%% START HERE
%%%%%%%%%%%%%%%%%% title page information %%%%%%%%%%%%%%%%%%
\title{Experimental generation of\\ amplitude squeezed vector beams}

\author{Vanessa Chille,$^{1,2,*}$ Stefan Berg-Johansen,$^{1,2}$ Marion Semmler,$^{1,2}$\newline Peter Banzer,$^{1,2}$ Andrea Aiello,$^{1}$ Gerd Leuchs,$^{1,2}$\newline Christoph Marquardt$^{1,2}$}

\address{\textsuperscript{1}Max Planck Institute for the Science of Light, G\"unter-Scharowsky-Str. 1/Bau 24, 91058
Erlangen, Germany\\
\textsuperscript{2}Institute for Optics, Information and Photonics, University of Erlangen-Nuremberg,
Staudtstr. 7/B2, 91058 Erlangen, Germany}

\email{$^*$vanessa.chille@mpl.mpg.de} %% email address is required



\begin{abstract} We present an experimental method for the generation of amplitude squeezed high-order vector beams. The light is modified twice by a spatial light modulator such that the vector beam is created by means of a collinear interferometric technique. A major advantage of this approach is that it avoids systematic losses, which are detrimental as they cause decoherence in continuous-variable quantum systems. The utilisation of a spatial light modulator (SLM) gives the flexibility to switch between arbitrary mode orders. The conversion efficiency with our setup is only limited by the efficiency of the SLM. We show the experimental generation of Laguerre-Gauss (LG) modes with radial indices up to 1 and azimuthal indices up to 3 with complex polarization structures and a quantum noise reduction up to -0.9dB$\pm$0.1dB. The corresponding polarization structures are studied in detail by measuring the spatial distribution of the Stokes parameters.\end{abstract}

 % REPLACE WITH CORRECT OCIS CODES FOR YOUR ARTICLE

%%%%%%%%%%%%%%%%%%%%%%% References %%%%%%%%%%%%%%%%%%%%%%%%%

%%%%%%%%%%%%%%%%%%%%%%%%%%  body  %%%%%%%%%%%%%%%%%%%%%%%%%%
\section{Introduction} \label{SecIntro}
Optical modes with complex structures have attracted increasing attention in recent years. The subject comprises very different issues due to the variety and versatility of higher-order modes, i.e. modes with complex intensity and phase patterns in their cross-sections.

Very prominent examples are light beams carrying orbital angular momentum (OAM), which is a direct consequence of a helical phase front of the beam \cite{Allan1992}. A set of solutions to the paraxial wave equation exhibiting OAM are so-called Laguerre-Gaussian (LG) beams. OAM has also been found to be an interesting additional degree of freedom for communication tasks that could help to meet increasing capacity demands \cite{NatPhotWang}.

Complex structures may not only be found in intensity and phase, but also in the polarization structure of optical beams. 
Such vector beams have interesting properties that can be utilized in diverse applications \cite{Zhan2009}. They may exhibit extraordinary structures that are non-separable and show intriguing properties similar to entangled quantum systems \cite{Aiello2015, Toeppel2014} that can be exploited for practical tasks, such as high-speed kinematic sensing \cite{BergJohansen2015}. They also have non-trivial focusing properties, allowing, for instance, to focus light to a tighter spot compared to homogeneously polarized modes \cite{Quabis, Dorn, Youngworth}. Tightly focused and polarization-tailored light beams have also proven to be versatile tools in nano-optics and plasmonics for studying the optical properties of individual nanostructures \cite{Mueller2007, Zuechner2008, Wozniak2015}. Moreover, very efficient light-atom coupling can be realised by shaping the input field into a radially polarized mode \cite{Sondermann}. When quadrature squeezed, vector beams represent a natural system for the study of hybrid quantum entanglement between the polarization and transverse spatial degrees of freedom \cite{CGabriel}. Higher-order vector beams may also serve as a building-block to perform interesting quantum computational tasks. An implementation of a cluster state generation using vector beams that may play essential roles in one-way quantum computation protocols is presented in \cite{ClusterStates}.

In general, it is surprising how little quantum features of vector beams have been studied up to now. Fundamental properties and aspects of the theoretical description in terms of quantum mechanics of these beams are still being discovered. Considering the vast range of possibilities for applications in this domain, it becomes clear that it is a worthwhile topic to explore.
Recently, we reported on a scheme for the experimental generation of squeezed spatial modes \,\cite{MarionsDraft}. That work was concerned with scalar Laguerre-Gaussian and Bessel beams only. Here, we want to go a step further and prepare amplitude squeezed vector beams. 
In a practical setting, several specific requirements have to be considered for the preparation of these modes. The option to switch flexibly between different kinds of high-order modes, or different orders of one type of mode, is highly desirable. 
An efficient technique using q-plates for the generation of vector beams \cite{QPlatesVecModes} unfortunately cannot offer this flexibility. The fiber-based technique used in \cite{CGabriel} is also not applicable as the squeezing was generated in a nonlinear medium supporting only a first-order subset of vector beams. A viable strategy is to perform the mode conversion after the nonlinear process. This technique has been realized previously \cite{CGabrielEPJD}, but a mode-converter was utilized which entails similar limitations as in \cite{CGabriel}. Converting the mode after generating the squeezing, makes us face another serious restriction regarding losses which must be kept low. A commonly used tools for performing arbitrary mode conversions are spatial light modulators (SLM). Given the currently still high intrinsic losses of such devices, restricting the losses to a low level is not a simple task. In \cite{Maurer}, an experimental setup is presented with which higher-order vector beams of a very high quality have been generated interferometrically. But even without considering the technical losses caused by the SLM, the system-inherent losses amount to 75\%. Also in the scheme presented in \cite{Han2013}, a good mode quality comes at the price of losses. These high losses make those interferometric methods unsuitable for preserving squeezing during the mode conversion process. Furthermore, for the sake of stability and simplicity of alignment, a setup is preferable in which an interferometer interfering spatially separated beams is avoided. 
In this paper, we present a scheme that meets all these requirements. It allows almost arbitrary vector beams to be generated from one setup. Apart from this flexibility, the particular double reflection technique involving a spatial light modulator avoids intrinsic losses as well as stability concerns as associated with a phase locking technique.
%%%%%%%%%%%%%%%%%%%%%%%%%%%%%%%%%%%%%%%%%%%%%%%%%%%%%%%%%%%%%%%%%%%
\section{Experimental Methods}
In the first step of the experimental procedure, the amplitude of a linearly polarized fundamental Gaussian mode is squeezed. In our experiment, this is achieved by exploiting the optical Kerr effect in a standard fiber \cite{Sagnac}. In a second step, the mode conversion from the fundamental to higher-order modes is performed via a double reflection technique involving a spatial light modulator. The resulting higher-order vector beams are analyzed in terms of their intensity profiles, the spatial distribution of the Stokes parameters and their level of amplitude squeezing. The corresponding experimental setup is sketched in Fig.~\ref{Setup}.
%%%%%%%%%%%%%%%%%%%%%%%%%%%%%%%%%%%%%%%%%%%%%%%%%%%%%%%%%%%%%%%%%%%
\subsection{Mode Preparation} \label{ModePrep}
As a light source, an ultrashort pulsed laser (Origami Onefive) producing 220\,fs pulses with a central wavelength of 1560\,nm is utilized. A Sagnac interferometer \cite{Sagnac} employing a polarization-maintaining single-moder fiber (FS PM 7811 by 3M) is used to amplitude squeeze the beam. Here, a 90:10 beam splitter is used to couple a strong and a weak light pulse into the two opposite ends of the fiber. Due to the third-order nonlinear Kerr effect, the strong pulse is quadrature squeezed. The two counter-propagating pulses interfere, and the quadrature squeezing is transformed into squeezing of the amplitude quadrature, if the input powers are adjusted in a suitable way. 
At the output of the Sagnac loop, we measure a quantum noise reduction in the amplitude squeezed beam of up to $-3.4\,\text{dB} \pm 0.1\,\text{dB}$ by direct detection at a radio frequency sideband at 10\,MHz (resolution bandwidth: 300\,kHz, video bandwidth: 3\,kHz ). The 461 data points of one trace measured by the spectrum analyzer have been divided in chunks of 15 data points. The error given for the squeezing is the standard deviation of the squeezing determined for these individual blocks.

\begin{figure}[htb]
\centering
\includegraphics[width = \textwidth]{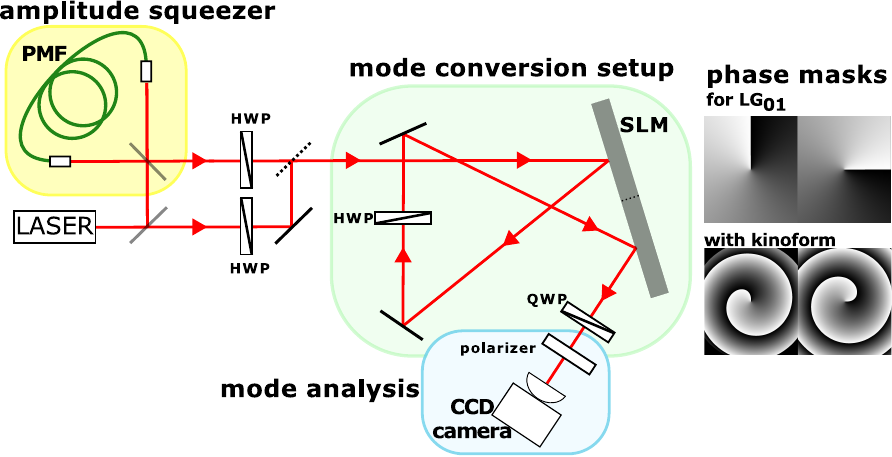}
\caption{Experimental setup for the generation of amplitude squeezed vector beams. The setup is divided into three main parts: The \textit{Sagnac loop} (yellow) that performs the quantum noise reduction in the amplitude quadrature, the \textit{mode conversion setup} (green) transforming the fundamental Gaussian mode into a higher-order Laguerre-Gauss mode and the \textit{mode analysis} setup (blue). 
The output of the laser is divided into two beams. One of them enters the Sagnac loop and is amplitude squeezed (see text in Sec.\,\ref{ModePrep} for details). The other one serves as a coherent reference beam which, by the help of a flip mirror, may be chosen to enter the mode conversion setup. We use it to evaluate the quantum noise reduction in front and behind the mode conversion setup via a direct detection scheme.
The mode conversion is performed via two individual spatial modulations of the vertically and horizontally polarized parts of the light beam in a double-reflection technique (see text for details). Examples for the phase masks for the two modulations are depicted here. In practice, a kinoform is added to the theoretical pattern to improve the mode quality. By using the quarter-wave plate after the SLM, the states of polarization are transformed from linear to circular. The higher-order vectorial mode arises from the superposition of these two individual modes propagating in the same beam path.}
\label{Setup}
\end{figure}

The prepared amplitude-squeezed fundamental Gaussian mode is then converted into the desired higher-order mode. For this mode conversion, the input beam undergoes two separate modulations which are performed subsequently. We use a reflective, phase-only liquid-crystal-based SLM (Pluto HOLOEYE, LCOS, 1920x1080 pixel) and split its display into two halves. On each half, we display individually tailored phase masks. The incoming beam is reflected from and modulated by the first pattern, before impinging on the second one (see Fig.\,\ref{Setup}). Due to the specific properties of the SLM, only the horizontally polarized part of the light is modulated. This polarization dependence is due to the elongated shape of the liquid crystal molecules and their birefringence.

We choose the state of polarization of the incoming light beam to be 45$^\circ$ relative to the SLM. The vertically polarized component of the light remains unaffected while the horizontally polarized component is transformed. After the first reflection, a half-wave plate rotates the states of polarization by 90$^\circ$, the modulated part is now vertically polarized, the unmodulated part horizontally polarized. The second reflection on the other half of the SLM thus causes the conversion of the so far unmodulated part of the light. The horizontally and vertically polarized components of the beam have now been reflected on different phase patterns which are tailored to produce the two different modes (see Fig.\,\ref{Setup}). We have chosen to work with LG modes. Thus, two copropagating LG-modes with opposite azimuthal indices (opposite phase charge) and opposite handedness (circular polarization) are generated. After passing these copropagating modes through a quarter wave-plate their interference results in the desired vector beams (see \cite{Maurer} for details). This method utilizing an LG-mode instead of Hermite-Gaussian (HG) mode basis turned out to be preferable. We achieved a better mode quality with a simpler phase pattern. This collinear common-path interferometer does not require phase-locking, as the prepared modes are co-propagating on the same beam path, they are naturally superimposed. The interference of these two modes leads to the desired polarization patterns within the beam profile, depending on the phase patterns displayed on the SLM. To improve the mode quality, we additionally added the phase pattern of a kinoform lens \cite{Kinoform}, and used combined patterns as illustrated in Fig.\,\ref{Setup}. The kinoform lens is a phase hologram that introduces a phase modulation equivalent to the one that would be caused by a real lens. Fig.\,\ref{FigLG01} and Fig.\,\ref{FigCCD} show the spatial images of the obtained modes.

Consistent with the expected losses in the conversion setup, the amount of squeezing is reduced. We measure a quantum noise reduction of up to $-0.9\,\text{dB} \pm 0.1\,\text{dB}$ in the converted modes. As in \cite{MarionsDraft}, no excess noise is added by the SLM at the measured sideband frequency. The overall losses in the setup amount to 64\,\%. It should be noted that these losses are not approach-inherent but solely due to the inefficiencies of some components, in particular of the SLM.

%%%%%%%%%%%%%%%%%%%%%%%%%%%%%%%%%%%%%%%%%%%%%%%%
\subsection{Mode Analysis} \label{AnalysisSetup}
We analyze the generated higher-order vector beams with regard to their mode structure and their quantum noise reduction in the amplitude quadrature. To this end, we measure the amplitude squeezing by direct detection with the help of a spectrum analyzer and a coherent beam serving as a reference level. The mode structure is studied by measuring the spatial intensity and polarization distributions in the beam cross-section. The total intensity profile is observed directly by using a CCD camera. A quick, yet quite precise check of the polarization structure is realized by inserting a rotatably mounted polarizer and taking images for different orientations of the polarization axis. A much more detailed analysis is the measurement of the spatial distribution of the Stokes parameters that we performed using a technique based on the method described in \cite{Schaefer}. With this technique, the beam profile after a rotatable quarter-wave plate and a polarizing beam splitter is determined by a CCD camera. From the images taken for different orientations of the quarter-wave plate, the Stokes parameters may be calculated using a Fourier algorithm described in \cite{Schaefer}. In this way, we obtain the polarization state for each pixel of the camera picture and, therefore, the spatial vectorial structure of the generated modes.
%%%%%%%%%%%%%%%%%%%%%%%%%%%%%%%%%%%%%%%%%%%%%%%
\section{Results}
As a proof-of-principle, we report in this manuscript the generation of Laguerre-Gauss modes with radial indices up to 1 and azimuthal indices up to 3, which have sophisticated polarization structures and show a quantum noise reduction in the amplitude of up to -0.9\,dB$\pm0.1$\,dB.
\begin{figure}[ht]
\centering
\begin{minipage}{\textwidth}\includegraphics[width = \textwidth]{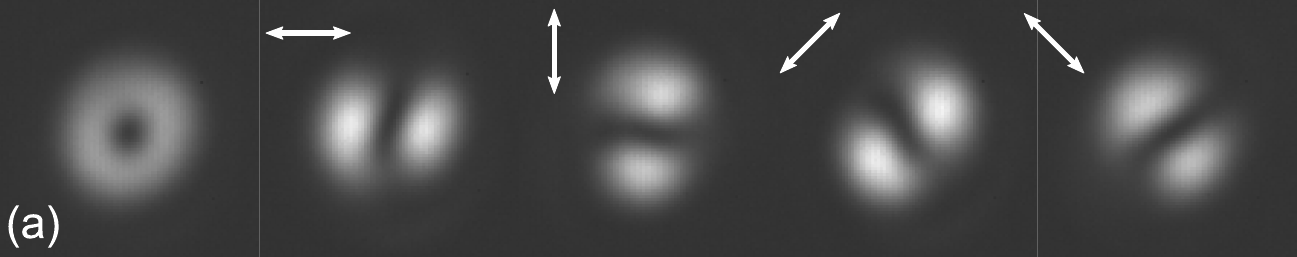}\end{minipage}
\hfill
\begin{minipage}{\textwidth}\includegraphics[width = \textwidth]{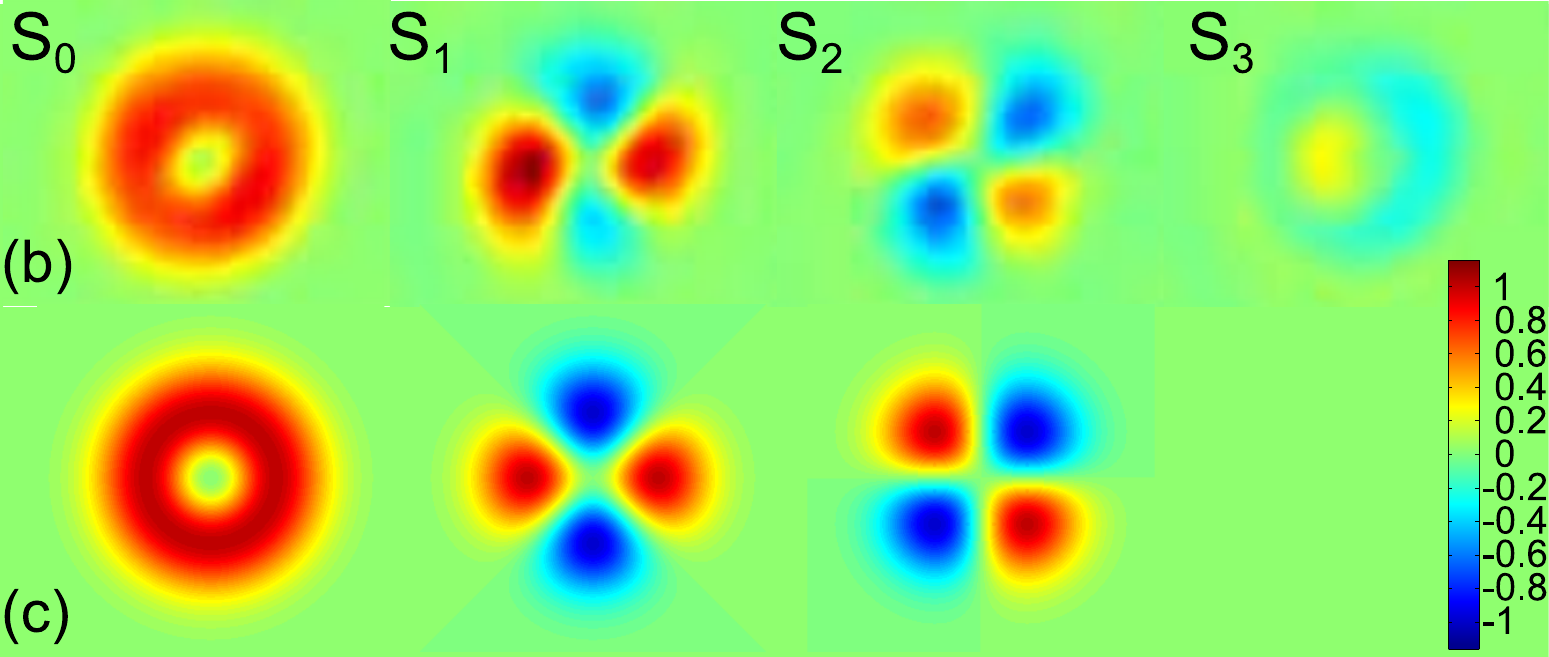}\end{minipage}
\caption{\textit{Analysis of the experimentally generated, radially polarized mode with the spatial structure of a LG$_{01}$ beam.} 
For the total intensity, the characteristic doughnut shaped pattern is observed, see \textit{(a)}. The following images represent the intensity distributions behind a polarizer orientated according to the white arrow. One can see a slight asymmetry due to the imperfections of the mode. The results of the spatial Stokes measurements are depicted in \textit{(b)}. Comparing them to the theoretical expectations shown in \textit{(c)}, we see in particular that there is a circularly polarized component that should not be there in the ideal case. The amplitude squeezing in the experimentally generated mode amounts to $-0.9\,\text{dB} \pm 0.1\,\text{dB}$ and is thus significant.}
\label{FigLG01}
\end{figure}

In Fig.\,\ref{FigLG01}, we illustrate our measured results and the theoretical expectation for a radially polarized LG$_{01}$ mode. In our nomenclature a LG$_{pl}$ mode is a Laguerre Gauss mode with the radial index $p$ and the azimuthal index $l$. Thus, here, the radial index is equal to 0, and the azimuthal index is 1. The first image in Fig.\,\ref{FigLG01}a) is the intensity pattern of the experimentally generated mode. The following images depict the intensity patterns for a polarizer being inserted in front of the camera, orientated in the direction indicated by the white arrows. Fig.\,\ref{FigLG01}b) shows the spatially resolved Stokes parameters of the experimental mode, Fig.\,\ref{FigLG01}c) depicts the theoretical expectations. We verified $-0.9\,\text{dB} \pm 0.1\,\text{dB}$ of amplitude squeezing in this mode. We encounter only small deviations, e.g. slightly imbalanced lobes for the Stokes parameter $S_1$ and a nonzero $S_3$ (for discussion see Sec.\,\ref{Sec_Discussion}).

\begin{figure}[ht]
\centering
\includegraphics[width = \textwidth]{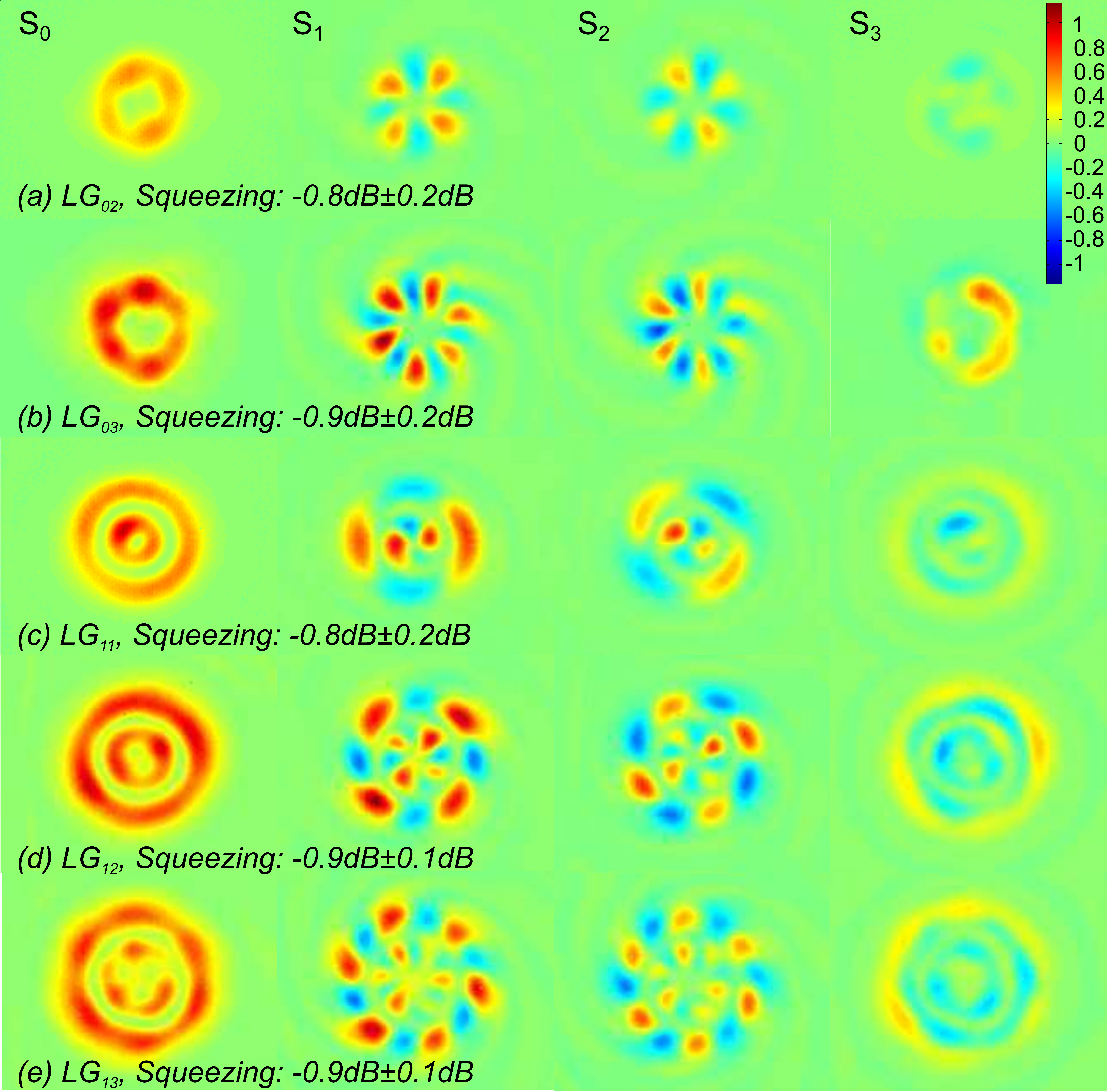}
\caption{\textit{Spatial distribution of the Stokes parameters of the experimentally generated higher-order modes with complex polarization structures} The Stokes parameters give a complete description of the modes and demonstrate that we achieve a good mode quality also for very sophisticated polarization structures. The error given for the amplitude squeezing is the statistical error.}
\label{FigStokes}
\end{figure}

Fig.\,\ref{FigStokes} shows our experimental results for LG$_{02}$, LG$_{03}$, LG$_{11}$, LG$_{12}$ and LG$_{13}$ modes with complex polarization patterns, the achieved quantum noise reduction is written directly below the images. The mode quality is high and the amplitude squeezing is significant for all of them. For the modes with two rings, i.e. for LG$_{11}$, LG$_{12}$ and LG$_{13}$, one can see that the polarization in the two rings is slightly twisted with respect to each other. This is in accordance with the spiral structures that can be noticed in the distribution of the Stokes observables (Sec.\,4). The effect increases with the complexity of the mode. Its cause is discussed in Sec.\,\ref{Sec_Discussion}. 

\begin{figure}[htb]
\centering
\includegraphics[width = \textwidth]{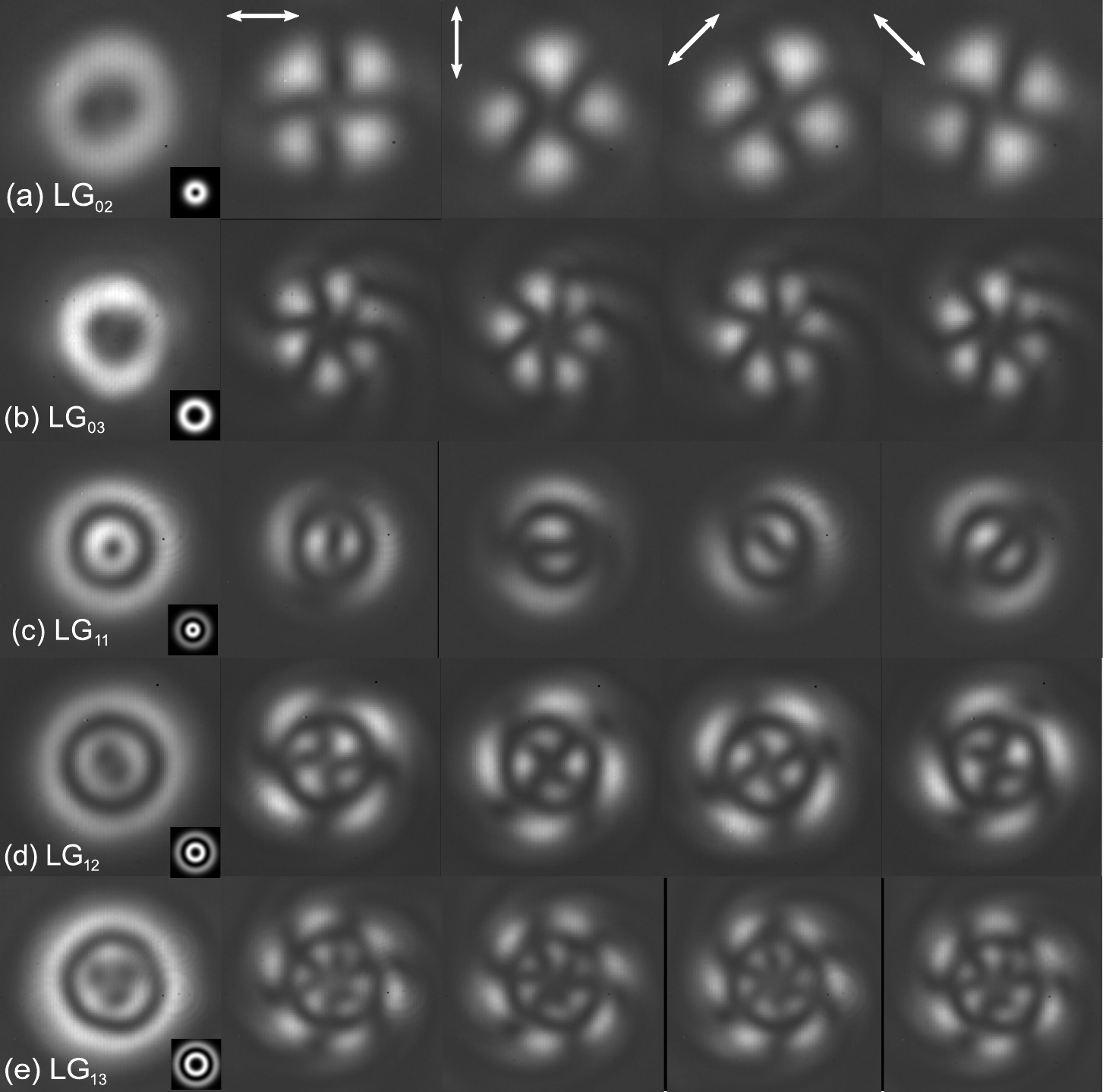}
\caption{\textit{Direct observation by means of a CCD camera} The first image in each row shows the intensity pattern of the mode. The following images are taken behind a rotatable polarizer, the orientation of which is indicated by the white arrows.}
\label{FigCCD}
\end{figure}

Apart from the spatial distributions of the Stokes parameters in Fig.\,\ref{FigStokes}, we also present the images from the direct observation of the modes with a CCD camera behind a rotatable polarizer in Fig.\,\ref{FigCCD}. They provide a depiction in a different manner that may give a more intuitive impression about some aspects of the mode quality. Both, Fig.\,\ref{FigStokes} and Fig\,\ref{FigCCD} demonstrate that our setup permits the flexible generation of different amplitude squeezed higher-order Laguerre-Gauss modes of good quality.
%%%%%%%%%%%%%%%%%%%%%%%%%%%%%%%%%%%%%%%%%%%%%%%%%%%%%%
\section{Discussion}\label{Sec_Discussion}
In the practical implementation of the mode preparation, special care has to be devoted to the overlap of the two orthogonally polarized modes stemming from the reflections on the two individual segments of the SLM. Since each polarization component interacts with the SLM in a different plane, the phase patterns imposed on each half must be carefully tweaked to ensure that both components emerge from the setup having the same width and divergence. The generation of the modes becomes more difficult the higher their order, for the simple reason that the higher-orders possess more complicated radial structures that need to overlap properly. Since the two polarization components originate from two modes propagating within one beam, the necessity of a phase locking technique is avoided altogether. We want to note that this advantage comes at a price: the mode quality is reduced by the imperfect diffraction efficiency of the SLM. The SLM modulates only a part of the impinging horizontally polarized light, the unmodulated part of the light is just reflected.

When a polarizer is inserted for the mode analysis, a particular linear polarization is chosen. In this way, one particular higher-order mode and the equally polarized part of the zeroth-order Gaussian mode is picked. The interference of these modes causes characteristic spiral structures. 
The appearance of these patterns cannot be avoided as the separation of the zeroth-order is not possible in this low-loss setup (see Sec.\,\ref{ModePrep}) and remain an undesired side effect especially for higher orders. The characteristic spirally curved patterns are particularly visible in Fig.\,\ref{FigCCD}b). Studying the Stokes parameters shows that circularly polarized components are present, resulting in non-zero $S_3$ components (see Fig.\,\ref{FigLG01} and Fig.\,\ref{FigStokes}). Our technique requires the presence of the unmodulated vertically polarized light and thus the remaining fundamental Gaussian mode cannot be separated by adding a blazed grating to the phase picture, for example, as it is very common in other experimental schemes. In \cite{Maurer}, this procedure can be applied and the mode quality is enhanced in this way. However, the system inherent losses render this approach not applicable for our purposes as this would considerably degrade the amount of squeezing. In other cases, superimposing a fundamental Gaussian mode with a higher-order mode is intentionally exploited to analyze the phase structure of the higher-order mode in question, see \cite{NatPhotWang, OpticsLettXu}. 
%%%%%%%%%%%%%%%%%%%%%%%%%%%%%%%%%%%%%%%%%%%%%%%%%%%%%
\section{Conclusion}
In this work, we present a flexible technique for the experimental generation of squeezed higher-order modes with nonuniform polarization patterns. We show that it is possible to minimize losses in a setup employing spatial light modulators. The amplitude squeezing initially generated by a nonlinear fiber-Sagnac interferometer was reduced during the mode conversion process due to inevitable losses, but remained significant in the higher-order Laguerre-Gauss modes that we studied, reaching values of up to $-0.9\,\text{dB} \pm 0.1\,\text{dB}$. The detailed analysis of our generated modes shows that we achieve a high mode quality. We discuss degradations of intensity profile of the modes, especially important at higher orders, due to a remaining fundamental mode.

A conceivable possibility to further improve the mode generation could be the combination of our method with the mode conversion technique described in \cite{Morizur}. The repeated reflection on a deformable mirror allows to generate modes with an extremely high mode quality. Unfortunately, the multiple reflections also cause a significant increase of losses such that a trade-off between mode quality and acceptable losses has to be found. Another strategy to enhance the mode quality could be to maintain our present scheme using an SLM, but further optimize the phase patterns displayed on the SLM \cite{Karimi2013}.

As a proof-of-principle, the results presented in this article can serve as a basis to study continuous-variable quantum states of arbitrary vector modes and their use in quantum processing and metrology.
\subsection{Acknowledgments} 
VC would like to thank Markus Sondermann for his support concerning the code for the Stokes measurements.
\end{document}